\documentclass[fleqn,10pt]{wlscirep}
\usepackage[utf8]{inputenc}
\usepackage[T1]{fontenc}
\usepackage{lineno}

\usepackage{caption}
\usepackage{subcaption}
\usepackage{makecell}
\usepackage{hhline}
\usepackage{multirow}

\title{A large-scale longitudinal structured dataset of the dark web cryptomarket Evolution (2014--2015)}

\author[1,*]{Hanjo D. Boekhout}
\author[2,3]{Arjan A.J. Blokland}
\author[1]{Frank W. Takes}
\affil[1]{Leiden University, Institute of Advanced Computer Science, Niels Bohrweg 1, 2333 CA Leiden, Netherlands}
\affil[2]{Leiden University, Institute of Criminal Law and Criminology, Steenschuur 25, 2311 ES Leiden, Netherlands}
\affil[3]{Netherlands Institute for the Study of Crime and Law Enforcement (NCSR), De Boelelaan 1077, 1081 HV Amsterdam, Netherlands}

\affil[*]{corresponding author(s): Hanjo D. Boekhout (h.d.boekhout@liacs.leidenuniv.nl)}

\begin{abstract}
Dark Web Marketplaces (DWM) facilitate the online trade of illicit goods.
Due to the illicit nature of these marketplaces, quality datasets are scarce and difficult to produce.
The Dark Net Market archives (2015) presented raw scraped source files crawled from a selection of DWMs, including Evolution.
Here, we present, specifically for the Evolution DWM, a structured dataset extracted from Dark Net Market archive data.
Uniquely, many of the data quality issues inherent to crawled data are resolved.
The dataset covers over 500 thousand forum posts and over 80 thousand listings, providing data on forums, topics, posts, forum users, market vendors, listings, and more.
Additionally, we present temporal weighted communication networks extracted from this data.
The presented dataset provides easy access to a high quality DWM dataset to facilitate the study of criminal behaviour and communication on such DWMs, which may provide a relevant source of knowledge for researchers across disciplines, from social science to law to network science.
\end{abstract}
\begin{document}

\flushbottom
\maketitle

\thispagestyle{empty}

\section*{Background \& Summary}
The online trade of illicit goods, from drugs and card information (credit/ID) to far more sinister goods and services like weapons and assassinations, has steadily grown since the launch of the first Dark Web Marketplace (DWM), Silk Road, in 2011~\cite{christin2013traveling}.
In 2020, for the first time the reported total trade volume, through DWMs, exceeded \$1.5 billion~\cite{chainreport}.
One explanation for this growth lies in the anonymity and reduced risk of violence, due to a lack of face-to-face contact in DWMs.
Anonymity is facilitated through browsers such as The Onion Router (Tor)~\cite{dingledine2004tor} and payment is typically made with crypto currencies such as Bitcoin~\cite{nakamoto2008bitcoin}.
Additionally, many DWMs provide services such as a moderated forum and escrow services to improve the trust between vendors and their customers.
Users can, for example, promote or review the goods on sale on the forum.
DWMs that provide these services are also referred to as \emph{cryptomarkets}~\cite{martin2014lost, shortis2020drug}.
Naturally, law enforcement has put continual effort in disrupting cryptomarkets~\cite{shortis2020drug} and researchers have endeavoured to better understand these markets and accompanying forums~\cite{van2017new, rhumorbarbe2016buying, li2021demystifying, booij2021get, hiramoto2023illicit}.

One complicating factor for research on cryptomarkets is obtaining data.
As cryptomarkets facilitate illegal trade they are naturally disinclined to share such information.
During the early years of DWMs an effort was made to scrape a large set of the active DWMs.
A comprehensive collection of these scrapes called the \emph{Darknet Market Archives} was published by Branwen et al.~\cite{dnmArchives} in 2015.
This collection contains raw source files scraped from 99 different DWMs and has been used in a plethora of research~\cite{rhumorbarbe2016buying, li2021demystifying, booij2021get, hiramoto2023illicit}.
A major downside of the Dark Net Market archives for researchers is that it consists of raw source files and relevant data must thus first be extracted before it can be analysed.
Such extraction is not (always) straightforward as the structure of source files changes and the fora and marketplaces evolve over time.
Furthermore, source files can be incomplete, include captchas, or simply be missing.
Ideally the effort of extracting, cleaning, and resolving inconsistencies in the data from the source files, would not need to be repeated by each individual researcher.
To assist with this, we present here a (near) complete and cleaned structured dataset for one of the cryptomarkets included in the Dark Net Market archives, \emph{Evolution}.

\begin{figure}[t]
  \centering
  \includegraphics[width=\textwidth]{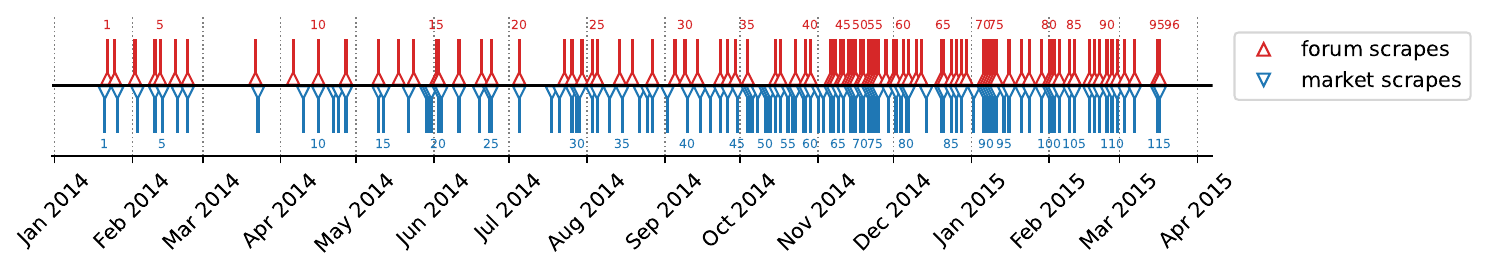}
  \caption{Timeline of scrapes of the Evolution forum and marketplace}
  \label{fig:timeline}
\end{figure}

Evolution was active from January 2014 until March 2015, when it closed due to an exit scam.
At the time, it was one of the most popular cryptomarkets~\cite{shortis2020drug}.
It formed a combination of a carding forum, where card information (e.g., credit/debit/ID/etc.) is traded, and an underground drug market; and was in essence the successor to the TCF carding forum~\cite{evolutionbackground}.
The raw source files obtained from the Dark Net Market archives consist of respectively 96 and 115 scrapes of Evolution's forum and marketplace (see Figure~\ref{fig:timeline}).
The Evolution forum consists of sequential posts in topics which are themselves categorized within fora that group topics by general subject.
From this we extracted information regarding each scrape, forum, topic, post, and user.
The Evolution marketplace revolves around the products for sale and their vendors.
Store, product category, listing, and vendor profile source pages provide varying types of information on product listings and vendors.
As such, we extracted information from the marketplace regarding each scrape, product category, listing, vendor, and feedback message.
Obviously, linking data from both the forum and market is important for research on how the forum communication impacts market performance and functioning.
Unfortunately, the source files provide no way to link forum user accounts to vendor accounts.
Therefore, we perform user matching based on usernames to provide a simple but relatively effective method of linking the two types of data.

Finally, with this paper we share monthly snapshots of an extracted weighted temporal communication network, which captures the indirect communication and shared interest of forum users.
These networks and the structured dataset were used to investigate to what extent (future) successful vendors on the Evolution cryptomarket could be predicted without knowledge of forum post content~\cite{boekhout2023early}.
For each month in the active period of Evolution, the set of the most successful vendors were determined in terms of both current and future sales.
We then investigated the predictive performance of several forum activity measures, directly computed from the structured dataset, and network measures, computed from the network snapshots.
We found that a particular combination of activity and network measures were able to reliably reduce the set of candidate users for law enforcement to investigate.
Additionally, we found that a significant proportion of the users with the highest forum activity and network centralities could be classified as vendors or other key players.
Furthermore, we found that these measures could serve as early warning signals for future vendor success.
The extracted dataset and communication networks may be of use for further studies of DWMs; and, the included communication network(s) constitute a rare relatively large sized weighted temporal network now available~\cite{boekhout_2023_10156522} open-source for use in network science research.
\section*{Methods}
\begin{figure}[t]
  \centering
  \includegraphics[width=\textwidth]{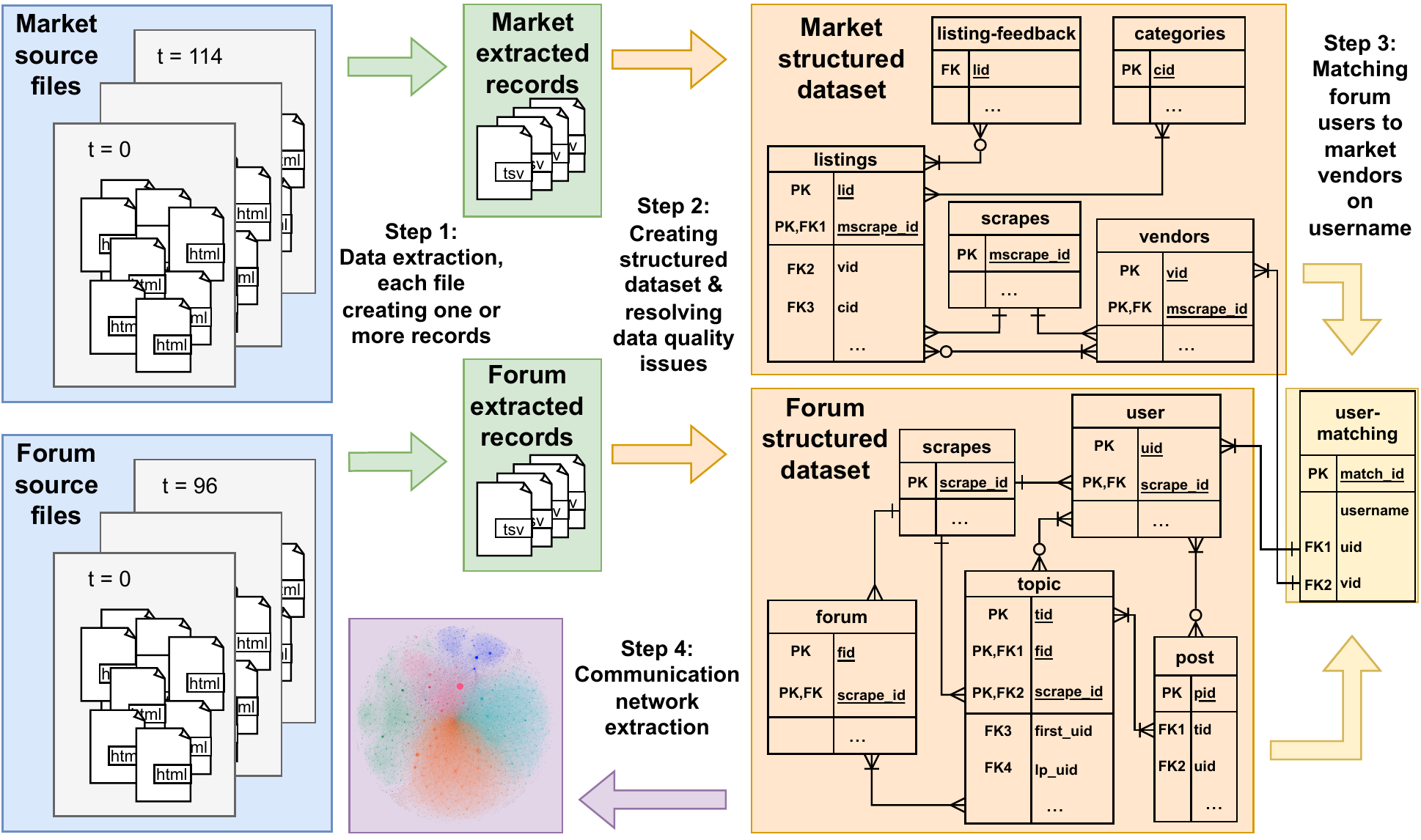}
  \caption{Overview of the four processing steps applied to our dataset}
  \label{fig:method-overview}
\end{figure}

The process of extracting our dataset from the raw source files of the DNM archives can be divided into four steps (as depicted in Figure~\ref{fig:method-overview}):
(1) extraction of all relevant data from forum and market raw source files, on a file by file basis;
(2) creating a structured dataset;
(3) matching forum users to market vendors;
and (4) extraction of weighted communication network snapshots.
\subsection*{Step 1: Extraction of forum and market data from source files}
In its raw state, the data on the cryptomarket Evolution retrieved from the Darknet Market Archives is split between forum and market data, which is subsequently split into various \emph{scrapes}.
Each scrape is the culmination of (several days worth of) capturing a great number of pages on the market or forum as mostly \emph{.html} files (generated by \emph{.php} files).
The relevant data is hidden/distributed among these dumps of many files.
Additionally, evolving versions of the same files may exist among the various scrapes.
Furthermore, errors may have occurred while capturing the files and there likely was a lack of access to the parts of the forum restricted to non-regular users.
As such, the data available in the captured files contains many of the common data quality issues like: data duplication, (temporal) data inconsistencies, missing and hidden data.
As our first step in identifying and resolving these issues, we extract all relevant data 
on a file by file basis, and store it in a tabular format.

The extraction process (for both forum and market) starts off by indexing the scrapes, which are organised in separate folders labelled by their date (see Figure~\ref{fig:timeline} for a timeline).
For each of the 96 forum scrapes we can encounter four types of files with relevant data:
(1) \emph{index} files (1 or 2 per scrape), which list the various fora, including descriptions, topic and post statistics, and cumulative statistics on the number of topics, posts, and users overall;
(2) \emph{viewforum} files (26,115 total), which describe a single page of topics that are part of a given forum (usually 30 topics);
(3) \emph{viewtopic} files ($\approx$5.2 million total), which describe a single page of posts that are part of a given topic (usually 25 posts);
and (4) \emph{profile} files (435,393 total), containing information on a given forum user.
From these files, we extract all available information regarding the various fora, topics, posts and users.

The Evolution marketplace provided various ways to browse the available listings: specific listing pages, store pages with listings sorted by vendor and category pages with listings sorted by product type.
Listing and vendor information was extracted from each of these types of pages.
We note that listing pages come in three formats, each providing some shared and some unique information.
First, only the \emph{generic format} provides a product description and available shipping destinations.
Second, the \emph{feedback format} provides a single page of feedback on the listing.
Finally, the \emph{return policy} format provides the return policy associated with the listing.
Additionally, we extracted vendor information from vendor profile pages as these provide additional information on the vendors themselves, irrespective of their listings.
We note that the vendor profile pages come in five formats: a \emph{generic format} and \emph{feedback format}, providing no additional unique information; a \emph{legacy sales format} detailing any (verified) sales this vendor might have had on past DWMs; a \emph{pgp format} providing the vendor's PGP-key; and, a \emph{return policy format} listing the return policy the vendor adheres to.
We ignored feedback provided by profile pages of the feedback format in favor of feedback obtained from listing pages.

In terms of missing data due to problems during the capture of pages, we found a combined 329 empty files, 91 source files indicating an error occurred, and 830 partial files.
Additionally, among the market files, we found 274 files with no useful information due to the scraper being logged out and 2,591 files with an Evolution market update and/or welcome message obscuring all relevant information.
Together this comes down to only 0.05\% of all source files.
Data duplication and inconsistencies occur naturally due to shared information between pages.
Additionally, temporal data inconsistencies between the scrapes occur as information gets updated and the forum continues to grow.
We discuss these data quality issues and how they were resolved in the next subsection.

We note that, some dates are indicated as ``Today'' or ``Tomorrow'' and that files in a scrape may have been gathered over multiple days.
Therefore, to determine the actual dates in such cases, we rely on the (last) modification date and time of the file.
We store this date and time as the \emph{retrieval time} along with the information extracted from each file.
Though not perfect, the retrieval time can serve as a proxy to determine the most `recent' information.
\subsection*{Step 2: Resolving data duplication and inconsistencies in extracted forum and market data}
At this stage, we have tabular data describing information as extracted from single files.
Next, want to merge and resolve this into structured datasets for both the forum and market, such that we have one set of unified information at either the scrape or dataset level.
Below, we discuss our attempts to resolve the various data quality issues during this process.
\subsubsection*{Forum}
We resolve data issues for five types of forum data: scrape, user, post, topic, and forum data.

First, for scrape data, we merely have to merge the indexed forum scrape dates with the scrape global statistics retrieved from index pages.
The global statistics indicate the forum's self-reported number of fora, topics, posts and users at that time.


Second, we resolve user data.
User data is retrieved from profile pages, but also from posts' poster and editor information and first and last post information of topics.
For all but five instances, the username appears immutable throughout.
However, on these five occasions the same \emph{uid} (user identifier) is associated with two different usernames.
For these rare cases a manual choice was made such that each \emph{uid} consistently has the same username attached.
On 38 occasions two (or three) \emph{uid}'s have the same username. Evidence from manual inspection shows that these are likely cases of a user being banned and subsequently creating a new account with the same username.
Although we could resolve this by combining these \emph{uid}s, we have chosen to leave them be such that this behaviour is not lost.

A number of inconsistencies might occur.
Data inconsistencies, at the level of individual scrapes, occurred for the user's title and number of posts.
For the former we rely on a manual ordering of importance: Administrator > Market Moderator > Forum Moderator > Moderator > Public Relations > Banned > Vendor > Resident Medical Expert > Troll > Member > Guest > Sports Referee > Sports Fan.
To resolve the latter, we take the maximum number of posts observed (within the scrape).

Temporal data inconsistencies, from one scrape to the next, occur for the last post and number of posts information.
The former observes apparent mismatches with the reported number of posts, where, for example, the last post information may change from one scrape to the next without their number of posts changing.
Such mismatches occur because the last post information can only be retrieved from the profile page of the user, while a post count is reported on both the profile page and alongside each individual post.
Since the pages with posts may have been retrieved after the profile pages, the maximum post count found can include posts beyond the last post reported on the profile page.
Additionally, because the posts list the number of posts at time of retrieval and not at the time of post placement, there is no way to tell if an apparent last post in our data is truly the last post.
As such, even when the number of posts of a user indicated by one of their posts is higher than that reported on the profile page, there is no reliable way to update the last post information.
Thus, these temporal inconsistencies remain unresolved and we simply store the last post information as presented on the profile page.

Additional temporal inconsistencies occur for the registration date of users.
These inconsistencies can be attributed to cases where the registration date was indicated with ``Today'' or ``Tomorrow'' for a given scrape.
All inconsistencies appear to have come about with a retrieval time shortly after midnight while the true date was the day before.
We consider it likely that this was caused due to a difference between the scraper's local time and the server time zone.
As such, the registration date inconsistencies were resolved to always be the earlier date.
However, we must note that due to a lack of observable inconsistencies some retrieval time faults may remain undetected .


Third, we resolve post data.
We store only a single entry for each post for the entire dataset, instead of for each scrape.
We do this to limit the size of the dataset.
Furthermore, since the posts themselves are timestamped, the amount of information this discards is limited.
The only information loss occurs for posts that were (repeatedly) edited, since we only store the final version of the post text and the details of the final edit (who and when).
For 65 topics ($0.13\%$) one (or more) apparent gaps between retrieved posts was found, with a cumulative size of 3,525 missing posts ($0.69\%$).
However, post deletion leads to temporal data inconsistencies w.r.t. the sequence id's of posts in a topic.
As such, we chose to recompute a \emph{seq\_id} based on all posts (whether deleted or not) that we were able to retrieve.
Thus, information on the existence of these gaps is lost in the structured dataset.
We report on the exact topics and size of the gaps in post data in Table~\ref{tab:missing-posts-distribution}.

Further temporal data inconsistencies occurred for post placement dates.
Again, all of these inconsistencies could be attributed to cases where the post placement date was indicated with ``Today'' or ``Tomorrow''.
The cause of the inconsistency differed from those observed for the registration date of users on only 52 occasions.
On these occasions, the retrieval time we found was exactly 7 days after the scrape date, an apparent impossibility.
The retrieval dates of these files were indeed found to be incorrect.
Similar to the registration date, post placement date inconsistencies were resolved to be the earlier date.
Note, that like for the registration date, faults may remain undetected due to a lack of observable inconsistencies.

When a post was edited multiple times, it is impossible to tell if an inconsistency of the date with the same username is due to a data inconsistency or due to multiple edits by that user.
As such, we simply store the last edit information retrieved, regardless of whether this may rely on a faulty date or not.
Finally, we note that the editor's \emph{uid} is not recorded on the source page.
Therefore, we use the previously resolved user data to retrieve the \emph{uid} matching the username.
Here, a \emph{uid} data inconsistency is resolved by taking the largest \emph{uid}.


Fourth, we resolve topic data.
Topic data was retrieved from pages listing the various topics in a forum and pages listing posts of the topic.
Since topics may be moved from one forum to another, separate entries are kept per scrape for each forum it was found to be associated with in that scrape.
Each entry indicates whether the topic has been moved.
Note that, for a given topic and scrape, we report the expected number of posts, the actual number of posts retrieved, and, based on the resolved post data, the number of posts with a placement date up to and including the scrape date retrieved overall.
This may seem redundant, but it is not uncommon for these values to differ from one another.
In fact, they complement each other, as together they provide a picture of what is expected and actually retrieved for a given scrape and what was retrieved overall so far for a given topic.

Data inconsistencies, at the level of individual scrapes, occurred for the topic title, expected number of posts, number of views, and last post information.
Topic title inconsistencies were resolved by choosing the title with the latest retrieval time, i.e., the most recent title.
Data inconsistencies for the expected number of posts and views were resolved to their maximum observed values.
For the last post information, we always resolved to the version associated with the `latest' post placement date.
Finally, any (temporal) inconsistencies for first and last post user information were conformed to previously resolved user data.


Fifth and last, we resolve forum data.
For a given forum and scrape, we report the expected number of topics, the actual number of topics retrieved, and, based on the resolved topic data, the number of topics associated with the forum for all scrapes up to and including this scrape.
Furthermore, we report the expected number of topics and, based on the resolved post data for the topics associated with the forum up to and including this scrape, the number of posts with a placement date up to and including the scrape date retrieved overall.
Similar as for topic data, these may appear redundant yet complement one another.

Temporal data inconsistencies occurred only for the expected number of topics, which were resolved to the maximum.
\subsubsection*{Market}
We resolve data issues for five different levels of market data: scrape, vendor, category, listing, and feedback data.


First, scrape data for the market has no observable data quality issues. Therefore, we directly copied the indexed scrape data into the structured dataset.


Second, we resolve vendor data.
Vendor data was retrieved from both listing and vendor profile pages.
Similar to the forum usernames, vendor usernames appear immutable throughout.
On only seven occasions, the same \emph{vid} (vendor identifier) is associated with two different usernames.
Again, a manual choice was made such that each vid is associated with a single username.
On fifteen occasions the same username was used for two different \emph{vid}'s.
Similar as for forum users this appears to be associated with vendors being banned; thus, again we have chosen to leave this be such that this behaviour is not lost.

Data inconsistencies, at the level of individual scrapes, occurred for vendors' rank, sales, positive/neutral/negative feedback statistics, approval rating, and the disabled flag.
Inconsistencies were resolved to the information with the latest retrieval time.
This ensured that the highest rank, sales, feedback statistics and the most recent approval rating and disabled status was kept.

Between scrapes 12 and 13 a change occurred in the ranking system (see Table~\ref{tab:ranking-systems} for details on the rankings).
Note that the \emph{Total Sales} ranges of the right ranking system, i.e., the latter system, are in practice only check after the \emph{Revenue} and \emph{Feedback} requirements are met.
Thus, the sales of a vendor with a lower rank can exceed the indicated range. Feedback refers to the \emph{approval\_rating}, i.e., the proportion of positive and negative feedback that is positive.
\begin{table}[t]
  \centering
  \caption{An overview of the ranking system of vendors until (left) and after (right) May 5th 2014, showing the ranks and their requirements.} \label{tab:ranking-systems}
  \begin{tabular}{|r|c||r|ccc|} \hline
    \multicolumn{2}{|c||}{ Until May 5th 2014} & \multicolumn{4}{c|}{From May 5th 2014} \\ \hline
    Rank & Total Sales & Rank & Total Sales & Revenue & Feedback  \\ \hline
    Freshman    & $\leq 4$    & Level 1 & 0--24     &     n/a &  n/a \\
    Sophomore   & $\leq 8$    & Level 2 & 25--99    &   1 BTC & 90\% \\
    Junior      & $\leq 16$   & Level 3 & 100--249  &  10 BTC & 90\% \\
    Senior      & $\leq 32$   & Level 4 & 250--499  &  50 BTC & 90\% \\
    Premium     & $\leq 64$   & Level 5 & 500+      & 100 BTC & 90\% \\ \cline{3-6}
    Advanced    & $\leq 128$  \\
    Expert      & $\leq 256$  \\
    Master      & $\leq 512$  \\
    Grandmaster & $\leq 1024$ \\
    Godlike     & $> 1024$    \\ \cline{1-2}
  \end{tabular}
\end{table}


Third, we resolve product category data.
Category data was retrieved from category pages and details the parental hierarchy and basic details of each category.
Although mostly static, the parental hierarchy experienced small changes over time, such as introducing intermediate levels.
We resolve the parental hierarchy to a single hierarchy with the greatest amount of detail, thus including intermediate levels.
Furthermore, over time two categories were renamed: Disassociatives $\to$ Dissociatives and Drug Paraphernalia $\to$ Paraphernalia.
We enforced the former change, as this was a correction.
However, the latter was due to the category becoming a child category of the Drugs category.
As such, we chose to keep the initial more detailed name.


Fourth, we resolve listing data.
Listing data was retrieved from store, category and listing pages.
Although each source provides a listing identifier (\emph{lid}) with each listing, we found this identifier to be unreliable.
The reason for this is that we found a large number of \emph{lid}s that were reported as one higher or lower than their true value.
Additionally, as listing titles changed often to represent either new products or slight changes to a product, resolving this issue becomes an extremely complex task.
As our study did not directly use the listing data, we chose not to resolve these issues definitively.
However, we did manually fix the egregious cases where an \emph{lid} would incorrectly have been associated with more than one vendor.
Additionally, we resolved, at the level of individual scrapes, titles that were a substring of one another, to the longest version.
Outside of these cases, we simply preserve data by adding an entry for each title option for a given scrape.

For individual scrape and title combinations, we find data inconsistencies for price, product class, shipment origin, and product category identifiers.
For the product category identifiers, we first exclude parent options using the resolved parental hierarchy such that we are left with the most specific category option(s).
For all other and the remaining product category identifier conflicts, the information with the latest retrieval time is kept, ensuring the most recent product information is stored.


Fifth and last, we resolve listing feedback data.
Listing feedback was provided near anonymously.
That is, outside of 27 instances, only the first and last letter (presumably) are reported of the usernames of those posting the feedback.
As such, our only resolution step is to remove duplicate data with the same \emph{lid}, username, date, and message.


We note that for both forum and market data, information on various entities may be incomplete due to missing source pages.
We discuss how these cases can be recognized in the structured dataset and report their frequency in the Technical Validation section.
\subsection*{Step 3: Forum user and market vendor matching}
We now have structured datasets for the forum and market data.
However, there is not yet an identifier to connect users in the forum and market datasets.
Therefore, we require a method to link forum and market users not reliant on the known identifiers.
Perhaps the simplest, but also the most reliable, method is to match forum users and market vendors based on their respective usernames.
After all, if vendors wish to promote their listings on the forum, posting under the same username creates name recognition.
Name recognition in turn has been linked to improved trust~\cite{chen2003interpreting} and market outcomes~\cite{huang2012brand} (e.g., more sales).
As such, vendors are naturally encouraged to use the same username on both the forum as the market.

For matching forum users and market vendors based on username, we create the \emph{user-matching} table.
This table reports a \emph{match\_id}, uniquely identifying usernames with successful matches, along with the \emph{username}, \emph{uid}, and \emph{vid} of the (non-)match.
Unmatched forum users and market vendors are also reported but without \emph{match\_id}.
We find that out of the 4,275 vendor usernames, 2,586 (i.e, 60.5\%) could be matched to a forum username.
In 35 cases, a single \emph{match\_id} is associated with multiple matches, due to the username being associated with multiple \emph{uid} and/or \emph{vid}.
\subsection*{Step 4: Network extraction}
As our final methodological step, we extract a communication network~\cite{fonhof2018characterizing,jo2023stage} $G = (V,E)$, with node set $V$ and edge set $E$, from the resolved structured dataset.
Each node in $u\in V$ represents a single distinct forum user $u$ that has at least one post in our data.
Note, we consider the uids with the same username as the same user.

Let $p_{k,u,i}$ indicate the $i^{th}$ post by the user represented by node $u$ in the $k^{th}$ topic;
let $\phi_{k,u,i}$ indicate the ordinal number designating the place in the ordered sequence of posts of $k^{th}$ topic, i.e., if $p_{k,u,i}$ is the $20^{th}$ post in $k^{th}$ topic then $\phi_{k,u,i} = 20$; finally,
let $t_{k,u,i}$ be the timestamp indicating on what day and at what time the post $p_{k,u,i}$ was placed.
A timestamped weighted directed edge $(u,v,t_{k,u,i},\omega_{k,u,i,v,j}) \in E$ is included for each pair of posts $p_{k,u,i}, p_{k,v,j}$ that satisfy the following conditions:
\begin{itemize}
  \item $u \neq v$, i.e., we do not include self-edges;
  \item $\phi_{k,u,i-1} < \phi_{k,v,j} < \phi_{k,u,i}$, i.e, $p_{k,u,i}$ is user $u$'s first post after $p_{k,v,j}$ was posted;
  \item $\phi_{u,i} - \phi_{v,j} \leq \delta_{\phi}$, i.e., the posts are at most $\delta_{\phi}$ posts apart; and
  \item $t_{u,i} - t_{v,j} \leq \delta_t$, i.e., the posts are at most $\delta_t$ time apart.
\end{itemize}
As such, each edge represents an (indirect) interaction between two users, but also a shared common interest due to posting in the same topic.
We add weights ($\omega_{k,u,i,v,j,}$) to each edge, to indicate how strong the connection between the two users is based on how much time passed between their respective posts.
The logic being that a faster, and thus more direct response equates to a stronger tie.
Let $\omega_{lower}$ be the minimum edge weight and $t_{lim}$ be the amount of time until the weight is reduced to this minimum.
Then the weight $\omega_{k,u,i,v,j}$ is computed using the following exponential weighting function:
\begin{equation}
  \omega_{k,u,i,v,j} = \omega_{lower} + (1 - \omega_{lower}) * \frac{\exp(3\frac{t_{lim} + t_{k,v,j} - t_{k,u,i}}{t_{lim}}) - 1}{e^3 - 1}.
\end{equation}

Finally, for every topic, for which we have post data, we add edges $(u, v, t_{k,u,i}, \omega_{first})$ between the user ($v$) that posted the initial post in the $k^{th}$ topic and every user ($u$) that posted in the topic afterwards (once for every post).
Here, $\omega_{first}$ is the same value for every edge.
We include these edges because each response in a topic is in essence also an indirect response to the initial post, thus a connection is established through shared interest.

\begin{figure}[t]
  \centering
  \includegraphics[width=0.4\textwidth]{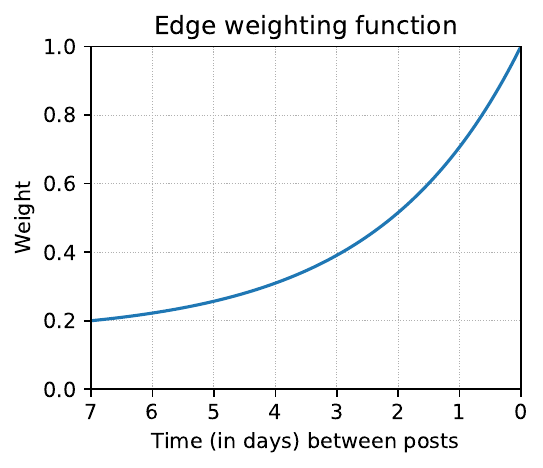}
  \caption{Exponential weighting for $\omega_{lower} = 0.2$ and  $t_{lim} = 7$ days.}
  \label{fig:weight-scheme}
\end{figure}

For the communication network included with this data descriptor, we used $\delta_{\phi} = 10$, $\delta_t = 1$ month, $\omega_{lower} = 0.2$, $t_{lim} = 7$ days, and $\omega_{first} = 0.5$.
Figure~\ref{fig:weight-scheme} plots the exponential weighting function for these values.
Different values for variables $\delta_{\phi}$, $\delta_t$, $\omega_{lower}$, $t_{lim}$, and $\omega_{first}$ can be given as parameters to the code published alongside this data descriptor to generate networks adhering to those conditions.

From the full communication network, monthly communication networks are formed.
Each monthly network includes only those edges that are formed from posts up to and including the final day of that month.
As such, each monthly network is a sub-graph of the subsequent monthly networks, i.e., the monthly networks grow over time.
A network is formed for each month that the Evolution cryptomarket was active, i.e., from January 2014 until March 2015.
\section*{Data Records}
The resolved structured dataset consists of five tables on forum data, five tables on market data, and one table on the results of the user matching.
Furthermore, the extracted monthly communication networks are presented in two types of tables, describing the nodes and (monthly) edges, respectively.
The full dataset is available on Zenodo by Boekhout et al.~\cite{boekhout_2023_10156522}.
All tables are given in tab separated format such that commas in post text, titles, and such did not need to be removed.
\subsection*{Forum}
The five tables describing forum data cover respectively: forum scrapes, fora, topics, posts and users.
\subsubsection*{forum/scrapes.tsv}
\begin{description}[leftmargin=!, labelindent=1cm, labelwidth=\widthof{\bfseries negative\_feedback}]
  \itemsep0em
  \item[\emph{scrape\_id}] forum scrape unique identifier;
  \item[\emph{scrape\_year}] year in which scrape was completed;
  \item[\emph{scrape\_month}] month in which scrape was completed;
  \item[\emph{scrape\_day}] day on which scrape was completed;
  \item[\emph{fora}] number of fora at the time of the scrape as reported by index page;
  \item[\emph{topics}] number of topics at the time of the scrape as reported by index page;
  \item[\emph{posts}] number of posts at the time of the scrape as reported by index page;
  \item[\emph{users}] number of users at the time of the scrape as reported by index page.
\end{description}
\subsubsection*{forum/forum.tsv}
\begin{description}[leftmargin=!, labelindent=1cm, labelwidth=\widthof{\bfseries negative\_feedback}]
  \itemsep0em
  \item[\emph{fid}] forum unique identifier;
  \item[\emph{scrape\_id}] forum scrape unique identifier (foreign key);
  \item[\emph{category}] forum category to which this forum belongs, values are \emph{Development, Discussion, Events, Information, Marketplace, Newbie Zone, Scams \& Trash} (with earlier version \emph{Trash}), and \emph{TCF};
  \item[\emph{title}] forum title;
  \item[\emph{description}] forum description;
  \item[\emph{pages}] number of pages listing topics in this forum, as implied by forum pages;
  \item[\emph{topics}] number of (expected) topics in this forum, as reported by statistics on index page or implied by forum pages;
  \item[\emph{topics\_visible}] number of unique topics actually found through forum pages alone;
  \item[\emph{topics\_found}] number of unique topics associated with this forum found through either forum or topic pages in all scrapes so far;
  \item[\emph{posts}] number of (expected) posts in this forum, as reported by index page;
  \item[\emph{posts\_found}] number of unique posts with a placement date up to and including the current scrape date in topics counted towards \emph{topics\_found}.
\end{description}
\subsubsection*{forum/topic.tsv}
\begin{description}[leftmargin=!, labelindent=1cm, labelwidth=\widthof{\bfseries negative\_feedback}]
  \itemsep0em
  \item[\emph{fid}] forum unique identifier (foreign key) to which the topic belongs or belonged (if \emph{moved} is True);
  \item[\emph{tid}] topic unique identifier;
  \item[\emph{first\_uid}] user unique identifier (foreign key) of the forum user that placed the first post of the topic;
  \item[\emph{scrape\_id}] forum scrape unique identifier (foreign key);
  \item[\emph{title}] topic title;
  \item[\emph{posts}] number of posts in this topic, as reported by statistics on forum pages or implied by topic pages;
  \item[\emph{posts\_visible}] number of unique posts actually found on topic pages;
  \item[\emph{posts\_found}] number of unique posts found in all scrapes so far, with placement dates up to and including the scrape date;
  \item[\emph{views}] number of views of this topic, as reported by statistics on forum pages;
  \item[\emph{lp\_uid}] user unique identifier (foreign key) of user who placed the last post, as reported on forum pages;
  \item[\emph{lp\_year}] year of last post, as reported on forum pages;
  \item[\emph{lp\_month}] month of last post, as reported on forum pages;
  \item[\emph{lp\_day}] day of last post, as reported on forum pages;
  \item[\emph{lp\_time}] time of last post, as reported on forum pages;
  \item[\emph{closed}] boolean flag indicating whether this topic was closed;
  \item[\emph{moved}] boolean flag indicating whether this topic was moved to another forum (fid).
\end{description}
\subsubsection*{forum/post.tsv}
\begin{description}[leftmargin=!, labelindent=1cm, labelwidth=\widthof{\bfseries negative\_feedback}]
  \itemsep0em
  \item[\emph{tid}] topic unique identifier (foreign key), indicating to which topic this post belongs;
  \item[\emph{pid}] post unique identifier;
  \item[\emph{seq\_id}] sequence id, indicating where in the order of posts of the topic this post belongs;
  \item[\emph{year}] year that the post was placed;
  \item[\emph{month}] month that the post was placed;
  \item[\emph{day}] day that the post was placed;
  \item[\emph{time}] time that the post was placed;
  \item[\emph{uid}] user unique identifier (foreign key) of the user who placed the post;
  \item[\emph{text}] the post's message text, as directly copied from source topic pages;
  \item[\emph{signature}] signature of the user who placed the post (most recent found for this post);
  \item[\emph{edit\_uid}] user unique identifier (foreign key) of the user who last edited the post;
  \item[\emph{edit\_year}] year of the last edit to the post;
  \item[\emph{edit\_month}] month of the last edit to the post;
  \item[\emph{edit\_day}] day of the last edit to the post;
  \item[\emph{edit\_time}] time of the last edit to the post.
\end{description}
\subsubsection*{forum/user.tsv}
\begin{description}[leftmargin=!, labelindent=1cm, labelwidth=\widthof{\bfseries negative\_feedback}]
  \itemsep0em
  \item[\emph{uid}] user unique identifier;
  \item[\emph{username}] username of user;
  \item[\emph{reg\_year}] registration year of user;
  \item[\emph{reg\_month}] registration month of user;
  \item[\emph{reg\_day}] registration day of user;
  \item[\emph{scrape\_id}] forum scrape unique identifier (foreign key);
  \item[\emph{title}] member title of user, indicating their role on the forum;
  \item[\emph{lp\_year}] year of last post placement, as reported on profile page;
  \item[\emph{lp\_month}] month of last post placement, as reported on profile page;
  \item[\emph{lp\_day}] day of last post placement, as reported on profile page;
  \item[\emph{lp\_time}] time of last post placement, as reported on profile page;
  \item[\emph{num\_posts}] number of posts by user to date, as reported on profile page or alongside the user's posts;
  \item[\emph{location}] self-reported location of forum user, highly unreliable.
\end{description}
\subsection*{Market}
The five tables describing market data cover respectively: market scrapes, product categories, listings, vendors, and listing feedback.
Note, neutral and negative feedback had to include a message, but positive feedback did not.
\subsubsection*{market/scrapes.tsv}
\begin{description}[leftmargin=!, labelindent=1cm, labelwidth=\widthof{\bfseries negative\_feedback}]
  \itemsep0em
  \item[\emph{mscrape\_id}] market scrape unique identifier;
  \item[\emph{scrape\_year}] year in which scrape was completed;
  \item[\emph{scrape\_month}] month in which scrape was completed;
  \item[\emph{scrape\_day}] day on which scrape was completed.
\end{description}
\subsubsection*{market/categories.tsv}
\begin{description}[leftmargin=!, labelindent=1cm, labelwidth=\widthof{\bfseries negative\_feedback}]
  \itemsep0em
  \item[\emph{cid}] product category unique identifier;
  \item[\emph{category}] name of category;
  \item[\emph{parent\_cid}] product category unique identifier of parent category of category.
\end{description}
\subsubsection*{market/listings.tsv}
\begin{description}[leftmargin=!, labelindent=1cm, labelwidth=\widthof{\bfseries negative\_feedback}]
  \itemsep0em
  \item[\emph{lid}] listing (unique) identifier;
  \item[\emph{vid}] vendor unique identifier (foreign key);
  \item[\emph{mscrape\_id}] market scrape unique identifier (foreign key);
  \item[\emph{title}] listing title, often includes basic description of product;
  \item[\emph{price}] price in BTC per unit of listing;
  \item[\emph{description}] description of product;
  \item[\emph{cid}] product category unique identifier (foreign key);
  \item[\emph{ships\_from}] indication of from what part of the world the product can be shipped;
  \item[\emph{ships\_to}] indication of to which parts of the world product can be shipped;
  \item[\emph{products\_class}] indication of type of products (either \emph{Digital} or \emph{Physical});
  \item[\emph{listing\_available}] indication (boolean) of whether product of listing remains available;
  \item[\emph{return\_policy}] return policy associated with the listing.
\end{description}
\subsubsection*{market/vendors.tsv}
\begin{description}[leftmargin=!, labelindent=1cm, labelwidth=\widthof{\bfseries negative\_feedback}]
  \itemsep0em
  \item[\emph{vid}] vendor unique identifier;
  \item[\emph{mscrape\_id}] market scrape unique identifier (foreign key);
  \item[\emph{username}] username of vendor;
  \item[\emph{rank}] rank of vendor (see Table~\ref{tab:ranking-systems} for details on the ranking systems);
  \item[\emph{sales}] number of sales vendor has reportedly completed so far (only available from market scrape 13 onwards);
  \item[\emph{approval\_rating}] rating computed as the proportion of positive and negative feedback received over all the vendor's listings that is positive;
  \item[\emph{positive\_feedback}] number of positive feedback responses the vendors has received on their listings;
  \item[\emph{neutral\_feedback}] number of neutral feedback responses the vendors has received on their listings;
  \item[\emph{negative\_feedback}] number of negative feedback responses the vendors has received on their listings;
  \item[\emph{legacy\_sales}] self-reported and `verified' sales of vendor on other Dark Web cryptomarket(s);
  \item[\emph{pgp\_key}] PGP-key used by vendor for secure communication with customers;
  \item[\emph{return\_policy}] return policy of vendor;
  \item[\emph{disabled}] indication (True or False) of whether the vendor account has been disabled.
\end{description}
\subsubsection*{market/listing-feedback.tsv}
\begin{description}[leftmargin=!, labelindent=1cm, labelwidth=\widthof{\bfseries negative\_feedback}]
  \itemsep0em
  \item[\emph{lid}] listing unique identifier (foreign key) to which feedback was placed;
  \item[\emph{username}] username of the one who posted the feedback, almost always hidden with only first and last letter visible;
  \item[\emph{year}] year that feedback was placed;
  \item[\emph{month}] month that feedback was placed;
  \item[\emph{day}] day that feedback was placed;
  \item[\emph{message}] feedback message.
\end{description}
\subsection*{User matching}
The matching of forum users and market vendors resulted in the following table.
\subsubsection*{forum-market/user-matching.tsv}
\begin{description}[leftmargin=!, labelindent=1cm, labelwidth=\widthof{\bfseries negative\_feedback}]
  \itemsep0em
  \item[\emph{match\_id}] successful username match unique identifier;
  \item[\emph{username}] username upon which forum user(s) and market vendor(s) were matched;
  \item[\emph{uid}] forum user unique identifier (foreign key);
  \item[\emph{vid}] market vendor unique identifier (foreign key).
\end{description}
\subsection*{Communication network}
The two types of tables describing the extracted communication network cover respectively the nodes and the (monthly) edges of the network.
Each edge table describes the edge list of a single monthly network and all edge tables follow the same format.
\subsubsection*{network/nodes.tsv}
\begin{description}[leftmargin=!, labelindent=1cm, labelwidth=\widthof{\bfseries negative\_feedback}]
  \itemsep0em
  \item[\emph{uid}] node identifier corresponding to forum user unique identifier (foreign key), lowest option if multiple options exist for the given username;
  \item[\emph{secondary\_uid}] second uid associated with the same username (foreign key);
  \item[\emph{tertiary\_uid}] third (and highest) uid associated with the same username (foreign key);
  \item[\emph{match\_id}] successful username match unique identifier (foreign key);
  \item[\emph{init\_year}] first year that the user placed a post on the forum;
  \item[\emph{init\_month}] first month (within \emph{init\_year}) that the user placed a post on the forum.
\end{description}
\subsubsection*{network/edges-2014-1.tsv, network/edges-2014-2.tsv, $\cdots$, network/edges-2015-3.tsv}
\begin{description}[leftmargin=!, labelindent=1cm, labelwidth=\widthof{\bfseries negative\_feedback}]
  \itemsep0em
  \item[\emph{Source}] source node forum user unique identifier (foreign key);
  \item[\emph{Target}] target node forum user unique identifier (foreign key);
  \item[\emph{Weight}] weight of edge, as computed based on time passed;
  \item[\emph{to\_first}] boolean indication whether this edge is based on linking to initial post, if True, \emph{Weight} $= 0.5$;
  \item[\emph{time\_diff}] elapsed time in-between the posts that generated this edge, expressed in seconds;
  \item[\emph{seq\_diff}] difference in \emph{seq\_id} between the posts that generated this edge;
  \item[\emph{timestamp}] timestamp of source post, expressed in seconds since January $1^{st}$ 2014;
  \item[\emph{tid}] topic unique identifier (foreign key), wherein the posts that generated this edge were placed.
\end{description}
\section*{Technical Validation}
\label{sect:tech-val}
In this section, we explore the technical quality of the dataset.
For a dataset extracted from scraped webpages over a period of time, the best way to assess the technical quality is to explore its completeness at various points in time and overall.
We report both on the completeness in terms of records and the frequency of missing fields on included records and their causes.
Additionally, we look at the amount of possible hidden data and discuss any data that was purposely not extracted and thus not included in the structured dataset.
Next, we discuss the reliability of certain `self-reported' data.
Finally, we have a look at the monthly communication networks that are included, exploring their growth over time as well as some basic statistics of the complete network.
\subsection*{Forum and market data completeness}
To investigate the completeness of the forum and market data we explore the completeness of the extracted dataset's records, look at the prevalence of empty fields within records, and consider the amount of hidden and missing data implied by entity identifier values.
\subsubsection*{Record completeness of forum and market data}
\begin{figure}[t]
  \centering
  \begin{subfigure}[b]{0.48\textwidth}
    \centering
    \includegraphics[width=\textwidth]{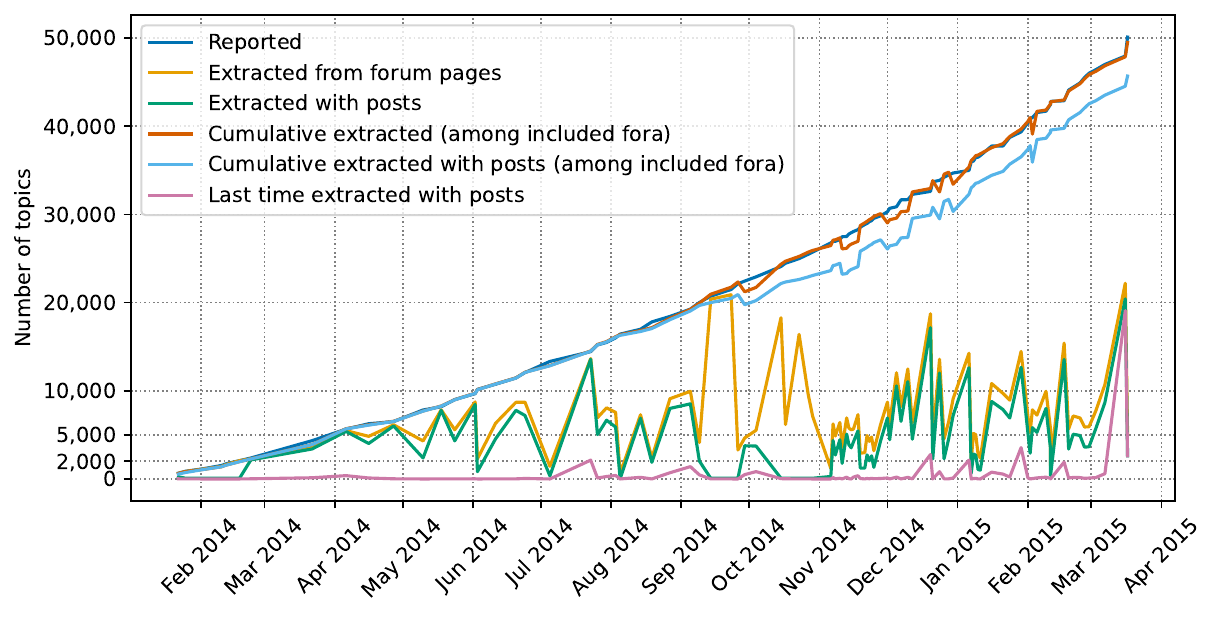}
    \caption{Forum topic statistics over time}
  \end{subfigure}
  ~
  \begin{subfigure}[b]{0.48\textwidth}
    \centering
    \includegraphics[width=\textwidth]{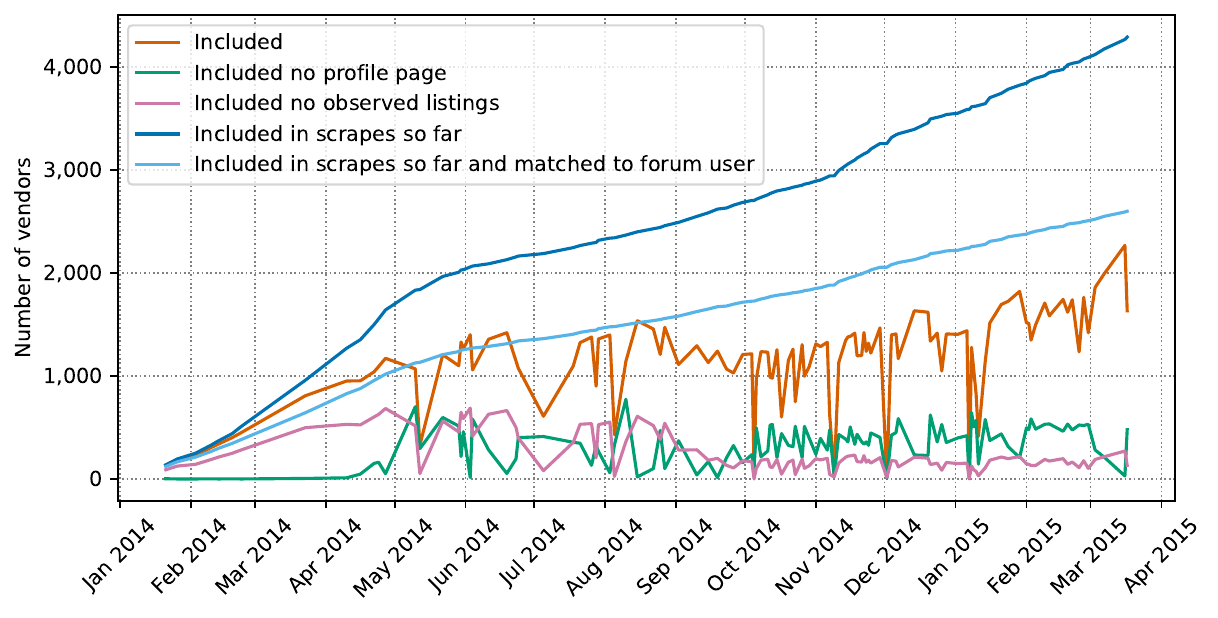}
    \caption{Market vendor statistics over time}
  \end{subfigure}
  \begin{subfigure}[b]{0.48\textwidth}
    \centering
    \includegraphics[width=\textwidth]{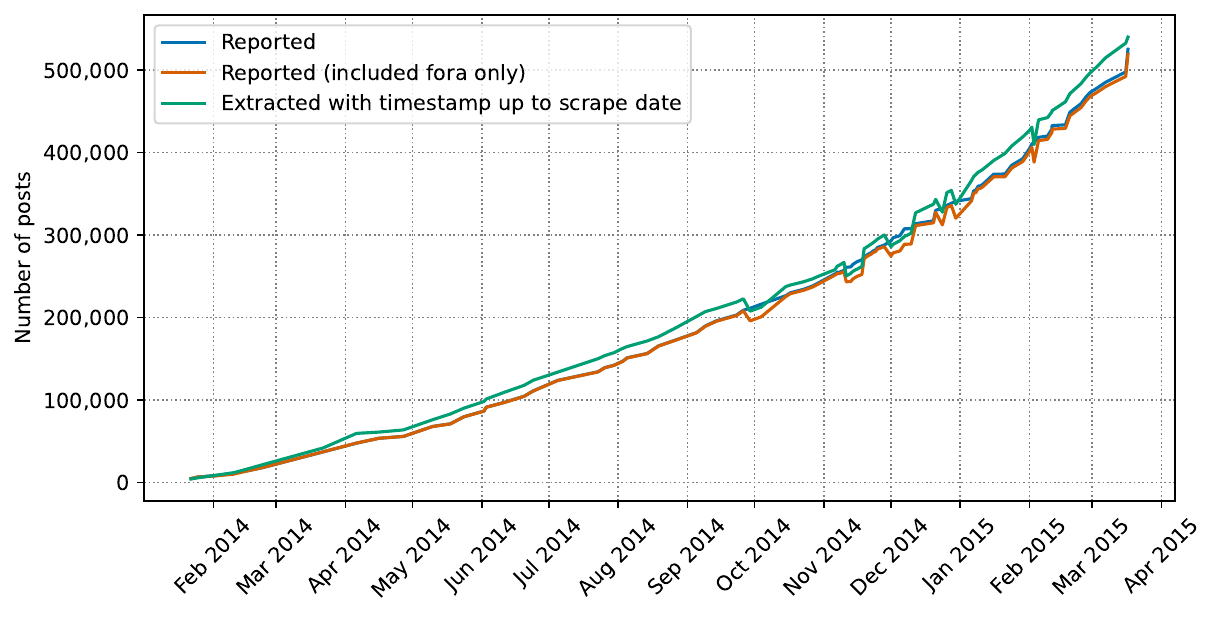}
    \caption{Forum post statistics over time}
  \end{subfigure}
  ~
  \begin{subfigure}[b]{0.48\textwidth}
    \centering
    \includegraphics[width=\textwidth]{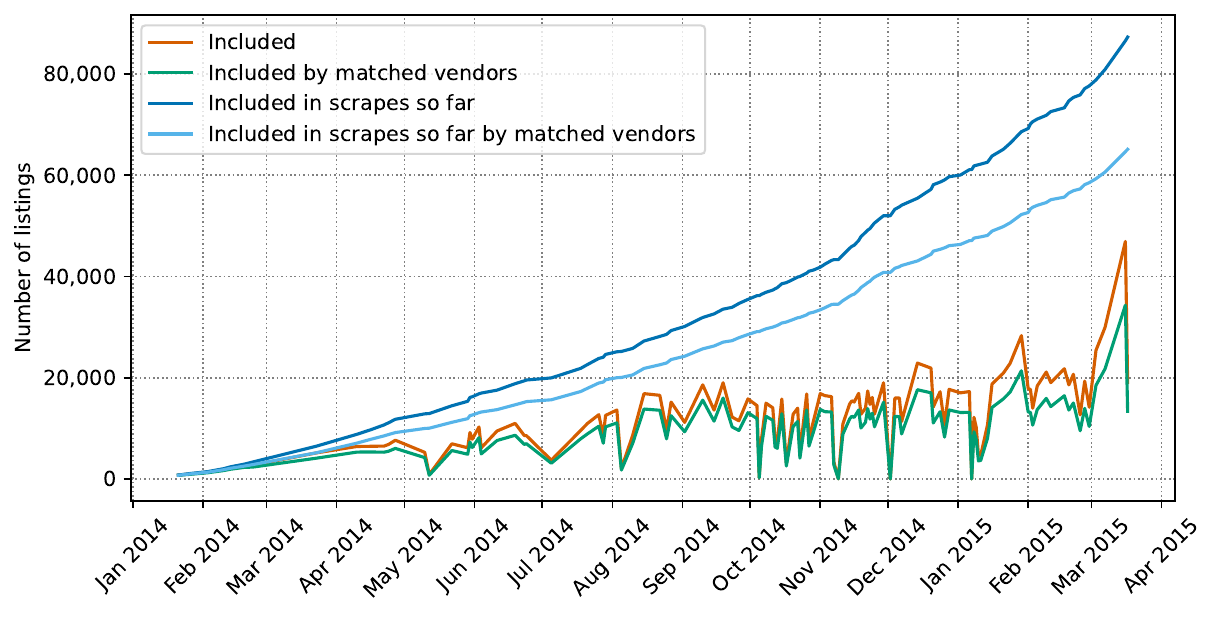}
    \caption{Market listing statistics over time}
  \end{subfigure}
  \begin{subfigure}[b]{0.48\textwidth}
    \centering
    \includegraphics[width=\textwidth]{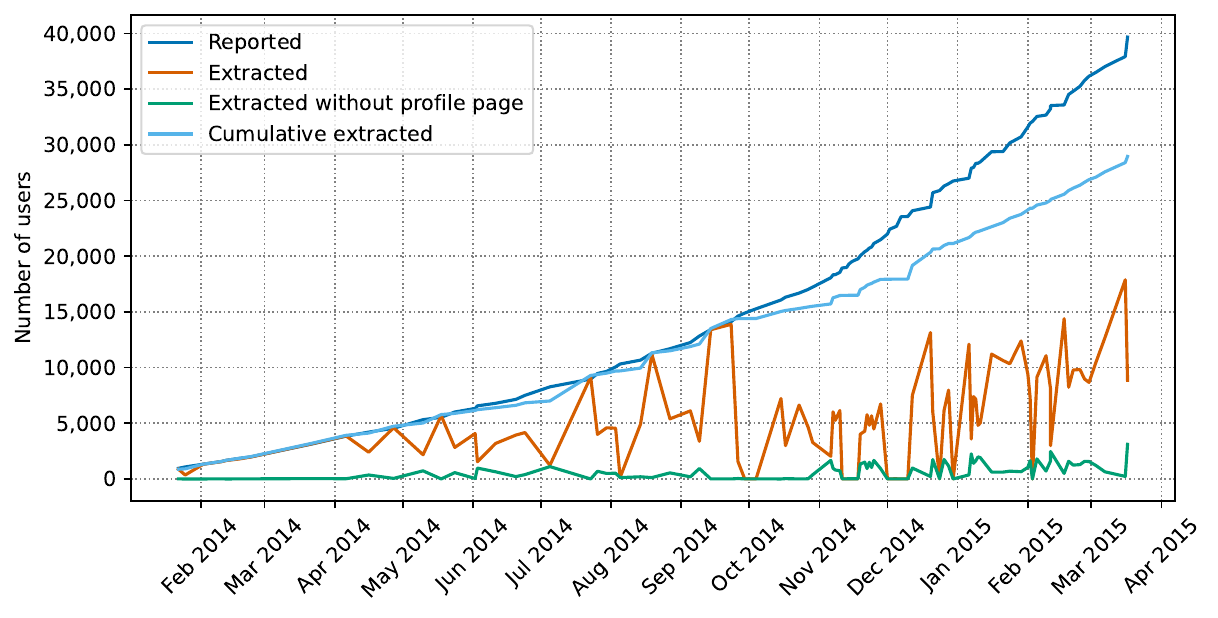}
    \caption{Forum user statistics over time}
  \end{subfigure}
  ~
  \begin{subfigure}[b]{0.48\textwidth}
    \centering
    \includegraphics[width=\textwidth]{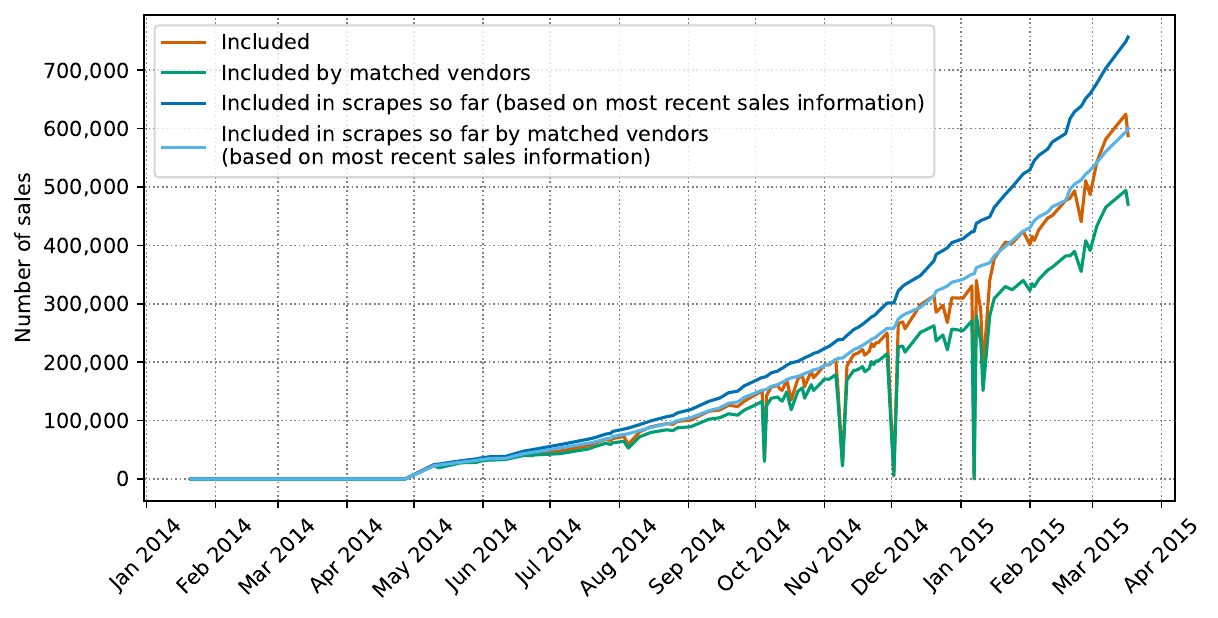}
    \caption{Market sales statistics over time}
  \end{subfigure}
  \caption{Statistics on the completeness of various elements of the dataset for each scrape, including statistics regarding the combined extracted data up to and including each given scrape.}
  \label{fig:completeness}
\end{figure}
First, we analyse the record completeness of forum topics, posts, and users and the market vendors, listings, and sales.
In order to argue about completeness we require some baseline truth to compare against.
For the forum data, we can compare the completeness of extracted data with the forum's self-reported statistics (obtained from the index pages).
Unfortunately, for the market no such statistics were reported.
Instead we compare the number of extracted vendors, listings, and sales overall with those associated with vendors that were matched to forum users.
This allows us to speculate whether non-matched vendors are likely due to a problem at extraction, or a fundamental lack of activity on the forum by these vendors.
Figure~\ref{fig:completeness} plots the completeness for each of the various elements over time, with statistics computed at the time of each scrape.
They combine plots on statistics for what was extracted for the specific scrape, with those displaying the cumulative amount extracted up to and including that scrape.

Figure~\ref{fig:completeness}a shows that the cumulative number of topics for which we extracted any data over time closely resembles the number of reported topics.
We notice that on a few occasions this cumulative number dips significantly while the reported number does not.
This phenomenon is caused by the ``Trash Can'' forum, which disappeared from the visible part of the forum during these periods of time while still being counted towards reported statistics.
Thus, we have data on (nearly) all topics on the part of the forum that is visible to all users.

We also consider the completeness of the extracted topics themselves, i.e., the number of topics for which posts were extracted at any point in time.
From Figure~\ref{fig:completeness}a we observe that, up to September 2014, posts were extracted for nearly all started topics.
However, during several scrapes in September and October 2014, almost no posts were extracted while the number of new topics continued its steady growth.
After this, the growth trends of the number of extracted topics and those for which posts were extracted are nearly identical.
As such, we conclude that it was primarily during this September/October period that we have missing posts data.
Specifically, between September 9th and November 7th, 69\% of the topics for which we were unable to extract posts were started and missed during scraping (note that another 12.5\% was missed on the final day of the cryptomarket).
In the end, we are left with 3,810 topics (7.7\% of total) with missing post data.
We note that there is no guarantee for all topics that if they include extracted post data, that this includes all post data.
After all, despite each scrape targeting the most recent (and oldest) topics for each forum, any given scrape only collected data on so many topics.
To study this, we also plotted how many topics had their posts extracted for the last time in any given scrape in Figure~\ref{fig:completeness}a.
This shows that nearly half of all topics had their posts extracted for the last time on the last two days of the existence of the cryptomarket.
The remaining observed spikes are closely related to spikes in overall extraction of topic post data for those scrapes.
As such, instead of implying that new post data was found for the last time at those times, the spikes more likely imply that they rediscover a set of topics that had already gone inactive at an earlier date for the last time.
For the remainder of the scrapes we observe a slow trickle of topics becoming inactive and no longer being rediscovered throughout.
Thus, outside of known missing posts due to gaps, we expect post records for nearly all topics, that have any post data at all, to be complete.

Figure~\ref{fig:completeness}c plots the reported and extracted number of posts over time.
However, as we only kept the final version of each post, we cannot plot for each scrape how many posts were extracted.
Instead, we plot the number of posts with a timestamp up to and including the scrape date for every scrape.
When comparing the actual number of extracted posts to the number of reported posts by the forum's own statistics, we find that we extracted more posts than were reported.
This is due to deleted posts not being counted towards official statistics, while still being included in our dataset.
During the September/October period, which resulted in 69\% of the topics with missing post data, the post surplus dropped by 9,940.
This translates to on average 3.8 posts per such topic.
Thus, the topics for which we have no post data at all were generally shorter topics that quickly became inactive.
Based on the post surplus at the end of the data and the decrease in surplus during the September/October period, we estimate an overall surplus of around 25 thousand posts ($\approx$ 4.8\% of reported posts).
As previously reported in the Methods section, gaps between posts were detected for 65 topics for a total of 3,525 presumed missing posts.
Table~\ref{tab:missing-posts-distribution} details for each topic (\emph{tid}) the observed gap sizes.
\begin{table}[!ht]
  \centering
  \caption{List of all observed gaps between posts that were subsequently excluded from the structured dataset. Each lists the topic in which the gap occurred, the \emph{pid} directly after the gap (i.e., the gap precedes this post), and the size of the gap.}
  \label{tab:missing-posts-distribution}
  \begin{tabular}{r|r|r||r|r|r||r|r|r}
    \emph{tid} & Before \emph{pid} & Gap size & \emph{tid} & Before \emph{pid} & Gap size & \emph{tid} & Before \emph{pid} & Gap size \\ \hline
    590   & 5874   & 1   & 9775  & 272486 & 1   & 29466 & 519090 & 48   \\
    616   & 6184   & 100 & 13698 & 393170 & 16  & 29604 & 540493 & 18   \\
    3116  & 547051 & 9   & 13698 & 424870 & 14  & 31243 & 396875 & 3    \\
    5029  & 536363 & 5   & 13698 & 444207 & 11  & 31656 & 551183 & 3    \\
    5029  & 553686 & 1   & 13698 & 553390 & 8   & 33679 & 306829 & 61   \\
    5275  & 547243 & 12  & 13698 & 555536 & 25  & 33865 & 304812 & 2    \\
    5275  & 558090 & 1   & 13698 & 557826 & 7   & 33865 & 304816 & 3    \\
    6549  & 229852 & 57  & 13961 & 199368 & 22  & 33865 & 307640 & 4    \\
    6549  & 231651 & 145 & 13961 & 203795 & 1   & 34282 & 311224 & 25   \\
    6549  & 245959 & 81  & 13961 & 261248 & 113 & 35324 & 555845 & 22   \\
    6549  & 306763 & 2   & 13961 & 325751 & 12  & 35506 & 395635 & 1    \\
    6549  & 312851 & 25  & 13961 & 399367 & 43  & 35506 & 551111 & 4    \\
    6549  & 492970 & 1   & 14267 & 132988 & 5   & 36363 & 343865 & 1    \\
    6549  & 503420 & 150 & 16751 & 543166 & 1   & 36573 & 548455 & 8    \\
    6549  & 554920 & 25  & 17259 & 269136 & 21  & 36831 & 519483 & 1    \\
    6549  & 558150 & 25  & 17437 & 519272 & 3   & 36831 & 519864 & 2    \\
    7354  & 557908 & 10  & 17967 & 557659 & 5   & 37091 & 528846 & 25   \\
    7448  & 191379 & 1   & 20772 & 198981 & 43  & 37615 & 351672 & 1    \\
    7448  & 205438 & 25  & 21107 & 467834 & 21  & 37615 & 358903 & 1    \\
    7448  & 228049 & 193 & 21897 & 552388 & 22  & 37750 & 524107 & 6    \\
    7448  & 231301 & 6   & 22305 & 546874 & 2   & 37750 & 535891 & 11   \\
    7448  & 271101 & 760 & 22410 & 523385 & 2   & 38449 & 537396 & 7    \\
    7448  & 496933 & 25  & 22410 & 531380 & 2   & 39031 & 522454 & 1    \\
    7448  & 498503 & 14  & 23010 & 219734 & 1   & 40089 & 553680 & 53   \\
    7448  & 557723 & 25  & 23010 & 549385 & 1   & 41410 & 557393 & 1    \\
    7906  & 542315 & 3   & 23337 & 255874 & 4   & 41804 & 540872 & 14   \\
    8746  & 549894 & 1   & 23337 & 258828 & 1   & 43961 & 557780 & 4    \\
    9543  & 199353 & 5   & 23635 & 554611 & 28  & 44789 & 551730 & 12   \\
    9543  & 374326 & 11  & 24517 & 512245 & 2   & 46401 & 554458 & 1    \\
    9543  & 392826 & 2   & 25810 & 239926 & 6   & 46403 & 537782 & 2    \\
    9543  & 429134 & 56  & 26245 & 231521 & 75  & 46847 & 545278 & 10   \\
    9543  & 442531 & 6   & 26551 & 245429 & 50  & 47391 & 557911 & 9    \\
    9543  & 501577 & 125 & 26910 & 245067 & 3   & 48606 & 518930 & 16   \\
    9543  & 554970 & 25  & 27273 & 550052 & 8   & 49823 & 478302 & 2    \\
    9543  & 558064 & 33  & 28178 & 336709 & 1   & 51493 & 556983 & 8    \\
    9775  & 229552 & 22  & 28178 & 427867 & 25  & 52862 & 552573 & 86   \\
    9775  & 237774 & 1   & 28178 & 517062 & 6   & 56556 & 559939 & 250  \\
    9775  & 265451 & 125 & 28178 & 558016 & 74  \\
  \end{tabular}
\end{table}

Completeness of forum user data is illustrated in Figure~\ref{fig:completeness}e.
User data is retrieved from profile pages, posts, post edit information, and topic information on the initial and final post.
Of these data sources, the profile pages are by far the most important for completeness of user data, as that is the only source to include users with zero posts.
We can easily see this reflected in the trends in Figure~\ref{fig:completeness}e.
After all, the number of extracted users only matches the reported number of users when all profile pages were included in a scrape.
The last time this occurred was near the end of September 2014, leading to a subsequent continuously widening gap between the number of extracted and reported users.
The final deficit in extracted users was 10,715 users (27\%).
Due to our general completeness of post data, we can surmise that the majority of users not included are likely those without any posts.

Despite that the majority of non-extracted forum users are likely without any posts, the size of the deficit may still have consequences.
After all, we could speculate that, if these users were included, a greater portion of vendors may have been matched to a forum user.
However, Figure~\ref{fig:completeness}b, which shows statistics on (matched) vendors, provides no evidence of this.
After all, it shows that the gap between the total number of vendors extracted and the number of them that could be matched to a forum user, grows steadily over time from the start.
If, as we previously speculated, our matching results were likely to have improved by including the missing forum users, we instead would have expected to see a relatively smaller gap right up till the end of September after which it would widen at a greater speed.
Thus, the missing forum users are unlikely to have posted nor to have been vendors.

Figure~\ref{fig:completeness}b shows that a significant number of vendors were not matched to a forum user.
In fact, at the end of our data only 60.6\% of vendors were matched.
Figures~\ref{fig:completeness}d,f show that this set of matched vendors is responsible for a much larger proportion of the overall listings and sales.
By the end of the data, the matched vendors cover respectively 79.3\% of listings and 74.6\% of all recorded sales.
Interestingly, each of the gaps between total and matched totals starts widening more starkly from November 2014 onwards.
Early that month, a joint international law enforcement operation dubbed ``Onymous''~\cite{shortis2020drug} lead to the closure of six cryptomarkets.
These closures aided in the growth of the two most popular cryptomarkets at the time, including Evolution.
The trends shown in Figures~\ref{fig:completeness}b,d,f imply that this event may also have impacted the behaviour of vendors and/or customers on the Evolution cryptomarket.
Vendors that emigrated to Evolution from one of the closed cryptomarkets may have felt less secure promoting their sales on the forum, leading to fewer new matches on average.
Additionally, customers that emigrated to Evolution may continue to purchase from their favourite vendors from previous cryptomarkets without these vendors needing to promote their listings.
This in turn could be one way to explain the increasing proportion of sales by non-matched vendors.
\subsubsection*{Field completeness of forum and market records}
So far we have discussed the completeness in terms of records.
However, even when records are included they are not always complete themselves, i.e., they may contain empty fields.
There are various reasons for fields of specific records to be empty.
In most cases, empty fields are due to missing source pages.
For example, we may have obtained data on a topic from topic pages, but not encountered them on forum pages.
In such a case, we would not have any information on the last post nor view statistics nor have any idea whether the topic had already been closed.
Other reasons for empty fields may simply be that the data did not exist to begin with.
Tables~\ref{tab:field-completeness-forum} and~\ref{tab:field-completeness-market} show, for the forum and market data respectively, the frequency of each type of occurrence of missing or otherwise distinctive field values, including the (likely) reasons for them.
\begin{table}[!ht]
  \centering
  \caption{Summary of reasons for empty fields (or specific values) in forum records, as well as their frequency w.r.t. all records of that type. Thus, lower frequency indicates greater completeness.} \label{tab:field-completeness-forum}
  \begin{tabular}{r|c|m{6cm}|r}
    Record type & Empty fields & Reason & Frequency (\% of records) \\ \hhline{=|=|=|=}

    \multirow{3}{*}{forum} & \makecell[l]{\emph{pages} \\ \emph{topics\_visible}} & Missing viewforum pages & 15 (0.76\%) \\ \cline{2-4}
                           & \makecell[l]{\emph{category} \\ \emph{description} \\ \emph{posts}} & Missing on index page but included through viewforum pages & 5 (0.25\%) \\ \cline{2-4}
                           & \makecell[l]{\emph{description}} & Information does not exist & 111 (5.64\%) \\ \hhline{=|=|=|=}
    \multirow{5}{*}{topic} & \makecell[l]{\emph{views} \\ \emph{lp\_uid} \\ \emph{lp\_year}, \emph{lp\_month} \\ \emph{lp\_day}, \emph{lp\_time} \\ \emph{closed} \\ \emph{moved} $\to$ False} & Missing viewforum pages (single scrape) & 2,189 (0.32\%) \\ \cline{2-4}
                           & \makecell[l]{Those listed above + \\ \emph{first\_uid}} & Missing viewforum pages for all scrapes for this topic & 165 (0.02\%) \\ \cline{2-4}
                           & \makecell[l]{\emph{views} \\ \emph{lp\_uid} \\ \emph{lp\_year}, \emph{lp\_month} \\ \emph{lp\_day}, \emph{lp\_time} \\ \emph{closed} $\to$ False \\ \emph{moved} $\to$ True} & Topic that was moved which was found on a viewforum page of its former forum but not on a viewforum page of the forum it was moved to. Viewtopic pages were found.
                           & 475 (0.07\%) \\ \cline{2-4}
                           & \makecell[l]{Those listed above + \\ \emph{posts}} & Same reasoning as above + missing viewtopic pages & 350 (0.05\%) \\ \cline{2-4}
                           & \makecell[l]{\emph{posts\_visible} $\to$ 0}  & Missing viewtopic pages & 239,398 (35.46\%) \\ \cline{2-4}
                           & \makecell[l]{Only: \\ \emph{closed}} & viewtopic pages were retrieved after topic was moved w.r.t. when viewforum pages were retrieved & 216 (0.03\%) \\ \cline{2-4}
                           & \makecell[l]{\emph{title}} & Information does not exist & 35 (0.01\%) \\ \hhline{=|=|=|=}
    \multirow{2}{*}{post} & \makecell[l]{\emph{signature}} & Information does not exist (poster had no signature) & 255,843 (49.74\%) \\ \cline{2-4}
                          & \makecell[l]{\emph{edit\_uid} \\ \emph{edit\_year}, \emph{edit\_month} \\ \emph{edit\_day}, \emph{edit\_time}} & Information does not exist (no edits occurred) & 455,254 (88.51\%) \\ \hhline{=|=|=|=}
    \multirow{3}{*}{user} & \makecell[l]{\emph{location}} & Information does not exist or missing profile page & 465,706 (93.52\%) \\ \cline{2-4}
                          & \makecell[l]{\emph{lp\_year}, \emph{lp\_month} \\ \emph{lp\_day}, \emph{lp\_time} \\ \emph{num\_posts} > 0 \\ \emph{location}} & Missing profile page \& posts by user found & 62,845 (12.62\%) \\ \cline{2-4}
                          & \makecell[l]{\emph{lp\_year}, \emph{lp\_month} \\ \emph{lp\_day}, \emph{lp\_time} \\ \emph{num\_posts} $\to$ 0} & No posts by user found & 48,351 (9.71\%) \\ \cline{2-4}
                          & \makecell[l]{All except: \\ \emph{uid} \\ \emph{username} \\ \emph{scrape\_id}} & Missing profile page \& no posts by user found & 105 (0.02\%) \\ \hline
  \end{tabular}
\end{table}
\begin{table}[!ht]
  \centering
  \caption{Summary of reasons for empty fields (or specific values) in market records, as well as their frequency w.r.t. all records of that type. Thus, lower frequency indicates greater completeness.}
  \label{tab:field-completeness-market}
  \begin{tabular}{r|c|m{6cm}|r}
    Record type & Empty fields & Reason & Frequency (\% of records) \\ \hhline{=|=|=|=}
    categories & \makecell[l]{\emph{parent\_cid}} & Already top level category itself & 13 (13.27\%) \\ \hhline{=|=|=|=}

    \multirow{4}{*}{listings} & \makecell[l]{\emph{ships\_from} \\ \emph{ships\_to} \\ \emph{product\_class} \\ \emph{listing\_available} \\ \emph{return\_policy}} & Missing all listing pages (\emph{mscrape\_id} <= 6) & 50 (0.00\%) \\ \cline{2-4}
                              & \makecell[l]{\emph{description} \\ \emph{ships\_from} \\ \emph{ships\_to} \\  \emph{listing\_available} \\ \emph{return\_policy}} & Missing all listing pages (\emph{mscrape\_id} > 6) & 614,628 (43.10\%) \\ \cline{2-4}
                              & \makecell[l]{\emph{description} \\ \emph{ships\_to} \\ \emph{return\_policy}} & Missing generic and return policy format listing pages & 109,605 (7.69\%) \\ \cline{2-4}
                              & \makecell[l]{\emph{description} \\ \emph{ships\_to}} & Missing only generic format listing page & 190,848 (13.38\%) \\ \cline{2-4}

                              & \makecell[l]{\emph{return\_policy}} & Missing only return policy format listing page & 508,435 (35.65\%) \\ \cline{2-4}
                              & \makecell[l]{(\emph{description} OR \\ \emph{ships\_to} OR \\ \emph{ships\_from}) \\ \emph{return\_policy}} & Partial generic format \& missing return policy format listing pages & 1 + 2 + 1 (0.00\%) \\ \hhline{=|=|=|=}

    \multirow{8}{*}{vendors} & \makecell[l]{\emph{sales}} & Always empty prior to \emph{mscrape\_id}$ = 13$ & 8,040 (6.01\%) \\ \cline{2-4}
                             & \makecell[l]{\emph{approval\_rating}} & No listings retrieved for this vendor OR listed as ``n/a'' & 27,513 (20.56\%) \\ \cline{2-4}
                             & \makecell[l]{\emph{positive\_feedback} \\ \emph{neutral\_feedback} \\ \emph{negative\_feedback} \\ \emph{legacy\_sales} \\ \emph{pgp\_key} \\ \emph{return\_policy} \\ \emph{disabled}} & Missing all profile pages & 36,122 (26.99\%) \\ \cline{2-4}

                             & \makecell[l]{\emph{legacy\_sales} \\ \emph{disabled} $\to$ False} & Missing legacy sales format profile page, but at least one profile page found & 87,224 (65.17\%) \\ \cline{2-4}
                             & \makecell[l]{\emph{pgp\_key} \\ \emph{disabled} $\to$ False} & Missing pgp format profile page, but at least one profile page found & 42,676 (31.89\%) \\ \cline{2-4}
                             & \makecell[l]{\emph{return\_policy} \\ \emph{disabled} $\to$ False} & Missing return policy format profile page, but at least one profile page found & 65,376 (48.85\%) \\ \cline{2-4}
                             & \makecell[l]{All except: \\ \emph{vid} \\ \emph{username} \\ \emph{mscrape\_id} \\ \emph{disabled} $\to$ True} & Vendor has been disabled & 440 (0.33\%) \\ \cline{2-4}
                             & \makecell[l]{As listed above, but \\ \emph{disabled} $\to$ False} & Vendor is scraper & 91 (0.07\%) \\ \cline{2-4}
                             & \makecell[l]{\emph{rank}} & Only an incomplete pgp format profile page & 1 (0.00\%) \\ \hhline{=|=|=|=}
    \multirow{2}{*}{listing\_feedback} & \makecell[l]{\emph{username}} & Unknown & 11 (0.00\%) \\ \cline{2-4}
                                       & \makecell[l]{\emph{message}} & Unknown & 42 (0.01\%) \\ \hline
  \end{tabular}
\end{table}
\subsubsection*{Hidden and not included data}
\begin{table}[ht]
  \centering
  \caption{Statistics on possible hidden or otherwise missing forum data, based on `unique' identifiers observed. The \emph{Maximum} reports the largest identifier found, while the \emph{Surplus reported last scrape} indicates how many more unique entities were reported to exist on the forum index page than were extracted for our dataset (\emph{Unique found}).} \label{tab:hidden}
  \begin{tabular}{r|rrr|r}
    Identifier & Maximum & Unique found & Surplus reported last scrape & Hidden (or missing) data (\% of maximum) \\ \hline
    \emph{fid} &      40 &      30 &       0 &      10 (25.0\%) \\
    \emph{tid} &  56,826 &  50,271 &     610 &   5,945 (10.5\%) \\
    \emph{pid} & 560,023 & 514,256 &       - &  45,767 (\ \ 8.2\%)\\
    \emph{uid} &  39,849 &  28,951 &  10,715 &     183 (\ \ 0.5\%) \\ \hline
  \end{tabular}
\end{table}
On nearly all fora, there exist parts that are only accessible to a select group as users, such as the moderators and administrators.
For our dataset, hidden data refers to this part of the forum that was likely inaccessible to those who gathered the data.
Although possible, no such hidden part of the marketplace is expected to exist, so we will not speculate on it here.
We can make an inference to the scope of hidden data on the forum by looking at the id's assigned to various elements.
For example, the highest observed \emph{fid} is 40, yet only thirty \emph{fid}s are recorded in our dataset.
This implies that there were (at least) ten hidden fora.

Table~\ref{tab:hidden} lists for each data element the maximum identifier, the number of different identifiers found, the surplus reported on the index page of the last scrape as compared to the number of unique found (when applicable), and the resulting possible amount of hidden data.
We note that the true amount of hidden data is likely to be less.
For example, we know that (at least) 3,525 posts are missing from topics with included posts and that for at least 3,810 topics no posts were extracted at all.
As such, far fewer posts were likely to have been associated with hidden fora than posited in Table~\ref{tab:hidden}.
Similarly, it is unclear if the reported number of topics on the index page includes the 1,287 closed topics we observed in our dataset.
Thus, Table~\ref{tab:hidden} clearly overestimates the true amount of hidden data and should not be taken as face value.

With respect to not included source data, the forum's image, wiki, and style files and the market's image and css files were not included in our dataset, but can easily be obtained from the original data source and were not suited to a structured dataset format.
The only remaining part of the source data that has not been included in the extracted dataset, but could have been, is the feedback on the vendor profile pages.
Instead only feedback from listing pages were obtained.
\subsection*{Data reliability}
Despite the (temporal) data inconsistencies discussed in the Methods section, in general we can consider any information that is not self-reported by users/vendors reliable.
Self-reported information however, may not be reliable.
Here we make the distinction between self-reported data that is considered a fact from the perspective of the forum/market, such as the post \emph{text} or listing \emph{description} and \emph{return\_policy}, and self-reported data that adds subjective value, such as a forum user's \emph{location}.
For the purpose of this section, we only consider the latter type as self-reported and potentially unreliable.

For the forum data, the only piece of self-reported data is the user's \emph{location}.
Looking at the actual contents of this field, for many records the \emph{location} information is indeed highly unreliable with entries such as: Everywhere, City of God, and Neptune.
For the market data, self-reported data consists of the listing's \emph{ships\_from} and vendor's \emph{legacy\_sales} and \emph{pgp\_key}.
The vendors' \emph{legacy\_sales} were (supposedly) verified by administrators and/or moderators, before they were added.
Furthermore, vendors' \emph{pgp\_key}s were crucial for the `safe' communication with the customer.
As such, the vendors' \emph{legacy\_sales} and \emph{pgp\_key}s are likely to be reliable.
For the market data, the \emph{ships\_from} field for listing records is the most likely to be inaccurate.
Like for the forum user's \emph{location}, \emph{ships\_from} information potentially exposes the vendor to discovery (by law enforcement).
Since the destination of a shipment is, form the customers perspective, not reliant on the origin of that shipment, vendors may be inclined not to report accurate information for \emph{ships\_from}.
Thus, outside of user/vendor location information, all other information is expected to be reliable.
\subsection*{Communication networks}
\begin{figure}[t]
  \centering
  \begin{subfigure}[b]{0.48\textwidth}
    \centering
    \includegraphics[width=\textwidth]{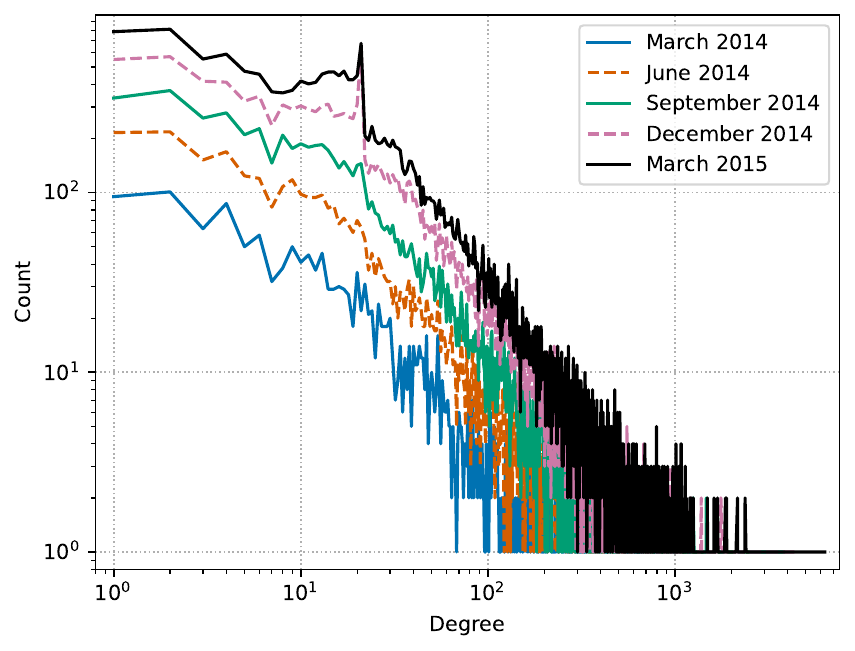}
    \caption{Degree distributions}
  \end{subfigure}
  ~
  \begin{subfigure}[b]{0.48\textwidth}
    \centering
    \includegraphics[width=\textwidth]{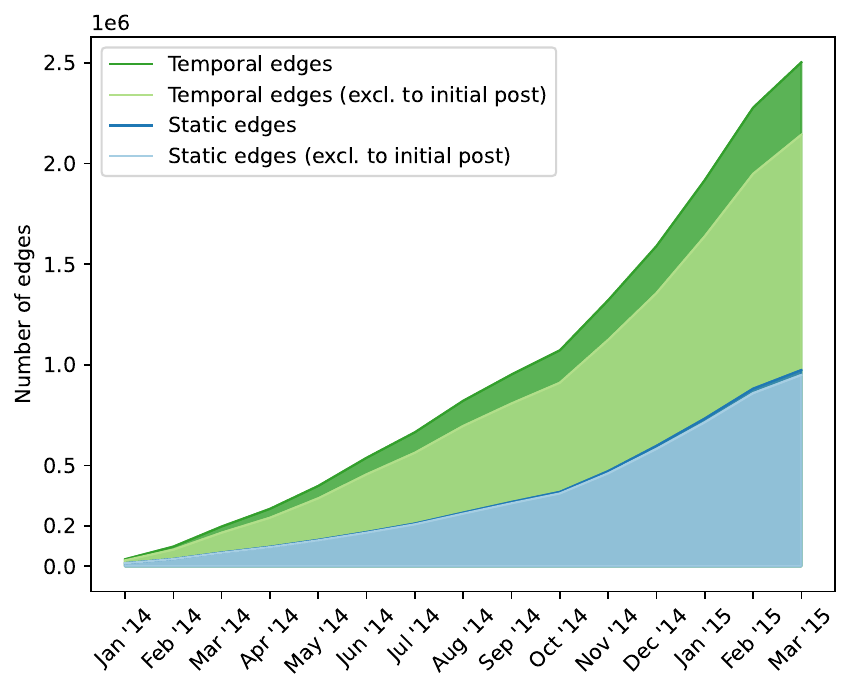}
    \caption{Communication network growth over time.}
  \end{subfigure}
  \caption{Degree distributions and communication network growth over time in terms of temporal and static edges. The darker parts indicate the edges formed solely by edges connecting to the initial posts of topics.}
  \label{fig:degdist-and-eot}
\end{figure}
\begin{table}[ht]
  \centering
  \caption{Basic statistics of full network (i.e., based on all data)}
  \label{tab:net-stats}
  \begin{tabular}{r|r}
    Measure & Value \\ \hline
    Nodes                                 & 21,986 \\
    Isolated nodes                        &    548 \\
    Temporal edges                        & 2,502,592 \\
    Static edges                          &   974,508 \\ \hline
    Number of weakly connected components (WCC)     & 14 \\
    Nodes in largest WCC (\% of non-isolated nodes) & 21,411 (99.874\%) \\
    Edges in largest WCC (\% of static edges)       & 974,490 (99.998\%) \\
    Number of strongly connected component (SCC)    & 2,151 \\
    Nodes in largest SCC (\% of non-isolated nodes) & 19,279 (89.929\%) \\
    Edges in largest SCC (\% of static edges)       & 967,348 (99.265\%) \\ \hline
    Density (excluding isolated nodes)          & 0.00212 \\
    Average clustering coefficient              & 0.598 \\
    Average clustering coefficient largest WCC  & 0.569 \\ \hline
    Diameter largest WCC (undirected, unweighted) & 7 \\
    Diameter largest WCC (directed, unweighted)   & 7 \\
    Diameter largest WCC (undirected, weighted)   & 8.28 \\
    Diameter largest WCC (directed, weighted)     & 17.70 \\ \hline

  \end{tabular}
\end{table}
In this subsection, we explore the basic properties of the full communication network based on all data.
Since we are dealing with a communication network, we only include those nodes with at least one post, thereby excluding the 6,925 nodes with no observed posts.
Some basic statistics for the full communication network are shown in Table~\ref{tab:net-stats}.
Here, isolated nodes are those users with at least one post but whose posts were the only ones in those topics, thereby establishing no links.
The average clustering coefficients and the sizes of the largest weakly and strongly connected components, tell us that this is a highly connected network.
Comparing with the networks in the KONECT~\cite{konect} database, we see that primarily some Human Contact and Human social networks have similar average clustering coefficients, but also an email communication network.
As such, the statistics observed in Table~\ref{tab:net-stats} gives us confidence that our extracted communication network is comparable to networks in the KONECT database, which are in common use.

The unweighted degree distributions for several iterations of the communication network are shown in Figure~\ref{fig:degdist-and-eot}a.
Over time, the distributions start showing spikes at degrees around twenty posts, usually corresponding to ten incoming and ten outgoing edges, that can be directly attributed to the manner in which the network was generated.
After all, given that each post looks forward and backwards at most ten posts to create respectively incoming and outgoing edges, it follows that many users with a single post in a relatively longer topic, will end up with an in- and out-degree of (almost) ten.
Furthermore, the longer topics responsible for this phenomenon were increasingly more likely to occur the older and more active the forum became.
For higher degree values, the degree distribution appears to follow a power law distribution not unlike many other real-world networks.

Figure~\ref{fig:degdist-and-eot}b shows the growth of the communication network over time in terms of edges.
We plotted this for both temporal and static edges and both including and excluding the `special' edges to the initial post.
Notably, the growth accelerates from November 2014 onwards.
As previously mentioned, this corresponds with the closure of several other cryptomarkets due to the joint international law enforcement operation ``Onymous''~\cite{shortis2020drug}.
Thus, we can clearly see that the closure of other cryptomarkets led to a marked increase in activity on the Evolution cryptomarket.

These preliminary statistics convince us that the data is useful for further studies of the Evolution cryptomarket and dark web market research, but also for general use in network science research.
\section*{Usage Notes}
The dataset comes in tab separated format and can be loaded using any standard data processing tool (python/R/etc.).
Figure~\ref{fig:method-overview} illustrates how the data tables can be linked through the various identifiers.
Note that unlike all other forum elements, posts are not stored on a scrape by scrape basis.
Posts relevant to a scrape should be determined based on their placement date and the scrape date.
Additionally, note that due to a small number of user identifiers that are associated with the same username (for both $uid$ and $vid$ as described in Methods), multiple entries with the same $match\_id$ can exist.
It is recommended for any analysis to combine data on users with the same username, since the data suggests these are in fact the same users who were banned but started a new account.

Since the dataset provides textual information on topic titles, post text, listing titles, and listing descriptions (to name a few), text analysis can be used, for example, to analyse drug overlap (the extent in which the same vendors offer different types of drugs), or to gather information on modus operandi and risk mitigation strategies.
For all such analysis we remind researchers that the textual information included in this dataset may still include html elements, which may have to be removed before performing text analysis tasks.
\section*{Code availability}
The code used to create, from the raw source files, the dataset described in this Data Descriptor is included as Supplementary Information and will be archived in a public repository upon acceptance. Included among these files is a README file with details on software versions used and instructions on how to reproduce our dataset using this code.
\section*{Acknowledgments}
The research that included the creation of this dataset, was funded by Politie \& Wetenschap as part of the `Criminaliteit en radicalisering voorspellen uit dark web forumnetwerken' project.
\section*{Author contributions statement}
H.B. extracted the dataset from the raw data, performed the data quality resolution, extracted the communication networks and wrote the manuscript.
All authors reviewed the manuscript.
\section*{Competing interests}
The author(s) declare no competing interests.

\bibliography{bibliography}

\end{document}